\documentclass[aps,prl,groupedaddress,twocolumn,floatfix]{revtex4}
\def\cuscn{$\kappa$-(BEDT-TTF)$_2$Cu(NCS)$_2$}
\newcommand{\lambdab}{{\mbox{\boldmath{$\lambda$}}}} 
\newcommand{\Gammam}{\underline{\underline{\mbox{\boldmath{$\Gamma$}}}}}
\newcommand{\Identity}{\underline{\underline{\bf I}}} 
\usepackage{graphicx}
\usepackage[]{amsmath}

\begin{document}


\title{An analytically solvable model of the effect of magnetic
  breakdown on angle-dependent
  magnetoresistance in a quasi-two-dimensional metal}


\author{Andrzej Nowojewski}
\altaffiliation{Current address: Trinity College, Cambridge, CB2 1TQ,
  United Kingdom}
\affiliation{Clarendon Laboratory, University of Oxford, Parks Road, Oxford OX1 3PU, United Kingdom}
\author{Paul A. Goddard}
\affiliation{Clarendon Laboratory, University of Oxford, Parks Road, Oxford OX1 3PU, United Kingdom}
\author{Stephen J. Blundell}
\altaffiliation{Corresponding author: s.blundell@physics.ox.ac.uk}
\affiliation{Clarendon Laboratory, University of Oxford, Parks Road,
  Oxford OX1 3PU, United Kingdom}
\email{s.blundell@physics.ox.ac.uk}



\date{\today}

\newcommand{\chem}[1]{\ensuremath{\mathrm{#1}}}

\begin{abstract}
We have developed an analytical model of angle-dependent
magnetoresistance oscillations (AMROs) in a quasi-two-dimensional metal in
which magnetic breakdown occurs.  The model takes account of all the
contributions from quasiparticles undergoing both magnetic breakdown
and Bragg reflection at each junction and allows extremely efficient
simulation of data which can be compared with recent experimental
results on the organic metal \cuscn. AMROs resulting from both closed
and open orbits emerge naturally at low field, and the model enables
the transition to
breakdown-AMROs with increasing field to be described in detail. 
\end{abstract}

\pacs{72.15.Gd, 71.18.+y, 71.20.Rv, 74.25.Jb}
\maketitle

\begin{figure}
\includegraphics[width=8.5cm]{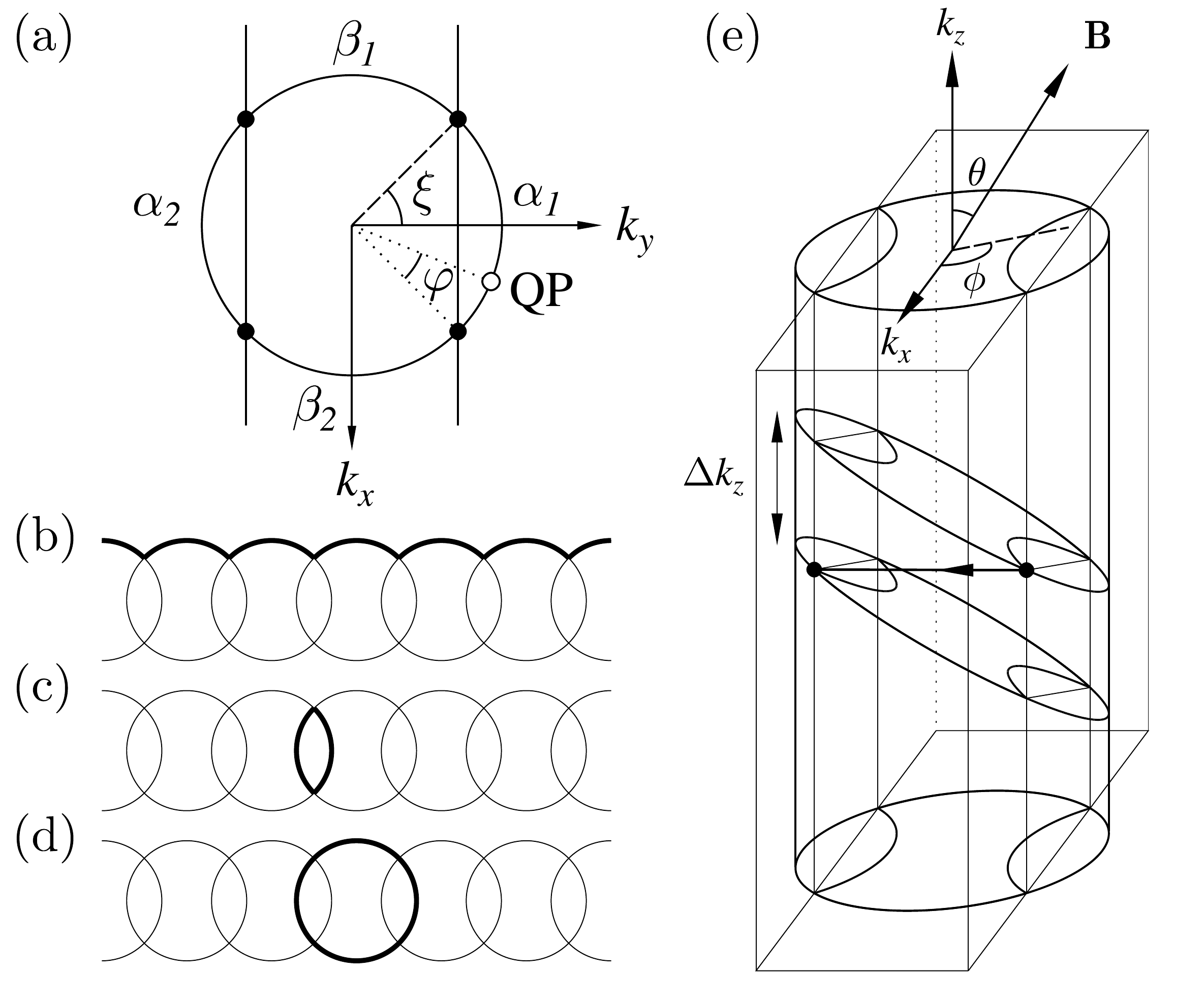}
\caption{(a) The Fermi surface (FS) in the $k_x$-$k_y$ plane showing the
  points where magnetic breakdown can occur which are at 
$(k_x,k_y)=(\pm k_{\rm F}\sin\xi,\pm k_{\rm F}\cos\xi)$ (these are
  called
MB junctions).  The
  azimuthal coordinate of a quasiparticle at the point labelled QP is 
$\varphi$.
(b) An open orbit (comprising the repeated traversal of the $\beta_1$
  section).  
(c) Closed orbit (comprising the repeated traversal of $\alpha_1$ and 
$\alpha_2$).
(d) Breakdown orbit (comprising $\alpha_1$-$\beta_1$-$\alpha_2$-$\beta_2$).  
(e) The magnetic field applied in a general direction leads to orbits
  which are on a cross-section perpendicular to ${\bf B}$.  Bragg
  reflection puts the quasiparticle on a different cross-section.
\label{fs}}
\end{figure}

The measurement of angle-dependent magnetoresistance oscillations
(AMROs) is a powerful technique in the determination of details of the
Fermi surfaces (FSs) in various reduced-dimensionality
metals~\cite{Kartsovnik2004,ruthenates,kawamura,Hussey2003}.  In many
cases the angle-dependence originates in correlations in the
time-dependent interplanar velocity of quasiparticles which traverse
the FS under the influence of the magnetic field $B$ and hence can be
efficiently simulated by integrating up such correlations for all
quasiparticle trajectories
\cite{lebed,Kartsovnik1988,Yamaji1989,Yagi1990,Danner1994,Blundell1996,kang}.
In high $B$, the additional effect of magnetic
breakdown (MB) can {\it substantially} complicate this picture.  This
effect occurs in the FSs of quasi-two-dimensional metals such as that
illustrated in Fig.~\ref{fs}(a) which is described by the dispersion
$E(k)=\hbar^2(k_x^2+k_y^2)/2m^*$ with effective mass $m^*$, Fermi wave
vector $k_{\rm F}$ and Brillouin zone edges at $k_y=\pm k_{\rm
F}\cos\xi$.  Because of the periodic potential, small gaps in the
dispersion open up at the Brillouin zone edge, splitting the FS into
distinct open and closed sections.  
Quasiparticles orbit around the FS
with constant $k_z$ when $B$ lies along the interlayer direction. In
very low $B$, because of Bragg reflection, 
only open orbits [Fig.~\ref{fs}(b)] and small
closed orbits [Fig.~\ref{fs}(c)] occur around the distinct sections of
the FS.  In high $B$, mixing between the states on the two FS
sections leads to MB at the four filled points shown in
Fig.~\ref{fs}(a) which we term MB junctions.  At these junctions a
quasiparticle ``tunnels'' in $k$-space between the FS
sections~\cite{Pippard1962}, resulting in a single large closed orbit
[Fig.~\ref{fs}(d)].

In fact for general values of the magnetic field there should be a
{\it superposition} of all the orbits in Fig.~\ref{fs}(b)--(d) as well
as many other intermediate possibilities in which MB occurs at some of
the MB junctions and Bragg reflection occurs at the others.  The
probability $p=\exp(-B_0/B)$ of MB at each MB junction is
parameterized by $B_0$, the characteristic breakdown
field~\cite{Pippard1962,Harrison1996,Shoenberg1984}.  For all
finite, non-zero values of $B$ (for which $0<p<1$) there is a
hierarchy of complex trajectories that must be summed to account for
all possible contributions to the conductivity in which MB either does
or does not occur at each MB junction.  If a quasiparticle crosses $N$
MB junctions, one has to consider $2^N$ possible trajectories with
their correct probabilistic weightings, and this complicates a direct
computation of AMROs since one has to sum over trajectories with
arbitrarily long path lengths and hence arbitarily large values of $N$.
In this Letter we describe a novel strategy to efficiently compute
AMROs in a model system exhibiting MB which includes explicitly all
these processes and we use it to show how different features in real
data may arise.  Our results are discussed in the context of recent
experimental work \cite{bamroexp} on the crystalline organic metal
\cuscn\ which demonstrated that, at high field, breakdown AMROs
(BAMROs) could be identified in experimental data resulting from
quasiparticles executing MB orbits, although until now
an adequate theoretical description has been lacking.

The Boltzmann equation gives the interlayer conductivity
$\sigma_{zz}=e^2\tau g(E_{\rm F})\langle v_z\bar{v_z}\rangle_{\rm FS}$
as an integral over the FS, where $\bar{v_z}=\int_0^\infty
\tau^{-1}{\rm e}^{-t/\tau}v_z[{\bf k}(t)]\,{\rm d}t$ and $g(E_{\rm
F})$ is the density of states at the Fermi energy.  Our model
considers the FS shown in Fig.~1(a) but includes a very weak
interlayer warping so that
$E(k)=\hbar^2(k_x^2+k_y^2)/2m^*-2t_\perp\cos k_zd_\perp$, where
$d_\perp$ is the interlayer spacing and the interlayer hopping
$t_\perp$ is small ($t_\perp\ll \hbar k_{\rm F}/d_\perp$). For
brevity, we will henceforth write wave vectors in units of
$d_\perp^{-1}$ and conductivity in units of $e^2 t_\perp^2 m^* d_\perp
/ \hbar^4 \pi^2 \omega$ so that they are dimensionless.  With ${\bf
B}=B(\sin\theta\cos\phi,\sin\theta\sin\phi,\cos\theta)$ quasiparticle
orbits lie in a plane perpendicular to ${\bf B}$ with angular
frequency given by $\omega=\omega_{\rm c}\cos\theta=eB\cos\theta/m^*$.
Neglecting the influence of $t_\perp$ on the quasiparticle motion
(which is only relevant for $\theta\approx 90^\circ$), an orbit can be
described by $k_z(t)=k_z^0+\eta\cos(\varphi-\phi+\frac{\pi}{2}-\xi)$
where $\varphi$ is the azimuthal position of the quasiparticle, given
at time $t$ by $\varphi=\varphi_0+\omega t$, and $\eta=k_{\rm
F}\tan\theta$.  For later convenience, we measure the azimuthal angle
$\varphi$ anticlockwise from the $\alpha_1$-$\beta_2$ MB junction [as
shown in Fig.~\ref{fs}(a)].  The interplanar velocity $v_z(t)$, which
is needed to compute $\sigma_{zz}$, can be written in our units as
$v_z(t)=\sin[k_z(t)]$.

When Bragg reflection occurs for a tilted orbit [see Fig.~\ref{fs}(e)]
the value of $k_z^0$ jumps by $\Delta k_z = 2\eta\sin\phi\cos\xi$
since only the $k_y$ value of the quasiparticle momentum changes and
hence the quasiparticle continues its orbit on a {\it different}
``slice'' of the Fermi surface. We can therefore write
$\sigma_{zz}={1\over\pi}\int_{-\pi}^\pi {\rm d}k_z^0 \int_0^{2\pi}
{\rm d}\varphi_0 \sin[k_z(0)] \int_{\varphi_0}^\infty {\rm d}\varphi
{\rm e}^{-(\varphi-\varphi_0)/\omega\tau} \sin[k_z(t)]$ where
\begin{equation}
 k_z(t)=  k_z^0+\mathsf{n}(\varphi)\Delta k_z
    + \eta \cos(\varphi-\phi+\frac{\pi}{2}-\xi),
\end{equation}
and where the term $\mathsf{n}(\varphi)\Delta k_{z}$ accounts for
jumps in the value of $k_z^0$ which occur during Bragg reflection and
we set $\mathsf{n}(\varphi_0)=0$.  The $k_{z}^0$ dependence can be
easily integrated out and we obtain
\begin{equation}
\sigma_{zz}=\int_{0}^{2\pi}d\varphi_0
E_{+}(\varphi_0)\int_{\varphi_0}^{\infty}d\varphi E_{-}(\varphi)e^{-{\rm i}
\mathsf{n}(\varphi)\Delta k_z},
\label{eformula}
\end{equation}
where the functions $E_{\pm}(x)$ are defined by
\begin{equation}
E_{\pm}(x)=e^{\pm{\rm i}
\eta\cos(x-\phi+\frac{\pi}{2}-\xi)\pm x/\omega\tau}.
\end{equation}
Eq.~(\ref{eformula}) yields a real expression for $\sigma_{zz}$ (one
can show straightforwardly that $\mbox{Im}\sigma_{zz}=0$).  However,
what makes Eq.~(\ref{eformula}) challenging to evaluate is that the
integrand changes depending on the path taken by the quasiparticle
which, at each MB junction of the orbit, can either undergo MB
tunneling (with probability $p\equiv {\rm e}^{-B_0/B\cos\theta}$) or
Bragg reflection (with probability $q=1-p$): this information is
encoded in the function $\mathsf{n}(\varphi)$ which remains constant
for MB but changes by $\pm 1$ for Bragg reflection.

A fruitful strategy is to follow separately the motion of particles
starting in the four different segments of the orbit, only finally
summing their contributions.  We therefore write Eq.~(\ref{eformula}) as
a scalar product of vectors
\begin{eqnarray}
 \sigma_{zz} = \lambdab_{+}\cdot
(\lambdab_{\textrm{init}}+{\bf x}_0-\lambdab_{-}),
\label{scalarproduct}
\end{eqnarray}
where $\lambdab_+$ takes care of summing up all the initial positions,
${\bf x}_0$ handles the MB junctions, $\lambdab_{\rm init}$ describes
the initial stage of the motion up to a MB junction and $\lambdab_-$
describes contributions between MB junctions.  In
Eq.~(\ref{scalarproduct}) we define
\begin{eqnarray}
\lambdab_{\pm}&=&\left(\begin{array}{c}\lambda^{\alpha_{1}}_{\pm}\\\lambda^{\beta_{1}}_{\pm}\\\lambda^{\alpha_{2}}_{\pm}\\\lambda^{\beta_{2}}_{\pm}\end{array}\right)=\left(\begin{array}{c}\int_{0}^{2\xi}d\varphi_0
E_{\pm}(\varphi_0)\\e^{\mp2\xi/\omega\tau}\int_{2\xi}^{\pi}d\varphi_0
E_{\pm}(\varphi_0)\\e^{\mp\pi/\omega\tau}\int_{\pi}^{\pi+2\xi}d\varphi_0
E_{\pm}(\varphi_0)\\e^{\mp(\pi+2\xi)/\omega\tau}\int_{\pi+2\xi}^{2\pi}d\varphi_0
E_{\pm}(\varphi_0)\end{array}\right), \nonumber \\
\lambdab_{\textrm{init}}&=&\left(\begin{array}{c}\int_{\varphi_0}^{2\xi}d\varphi
E_{-}(\varphi)\\e^{2\xi/\omega\tau}\int_{\varphi_0}^{\pi}d\varphi
E_{-}(\varphi)\\e^{\pi/\omega\tau}\int_{\varphi_0}^{\pi+2\xi}d\varphi
E_{-}(\varphi)\\e^{(\pi+2\xi)/\omega\tau}\int_{\varphi_0}^{2\pi}d\varphi
E_{-}(\varphi)\end{array}\right),
\label{lambda}
\end{eqnarray}
and ${\bf x}_0$ is a special case of the vector
\begin{equation}
{\bf x}_n = \left(\begin{array}{c}\alpha_1^n\\\beta_1^n\\\alpha_2^n\\\beta_2^n\end{array}\right),
\end{equation}
representing the contribution at the MB junctions on the $n$th slice of
the FS, where $n$ is one of the integer value taken by the function
$\mathsf{n}(\varphi)$.

\begin{figure*}
\includegraphics[width=17.5cm]{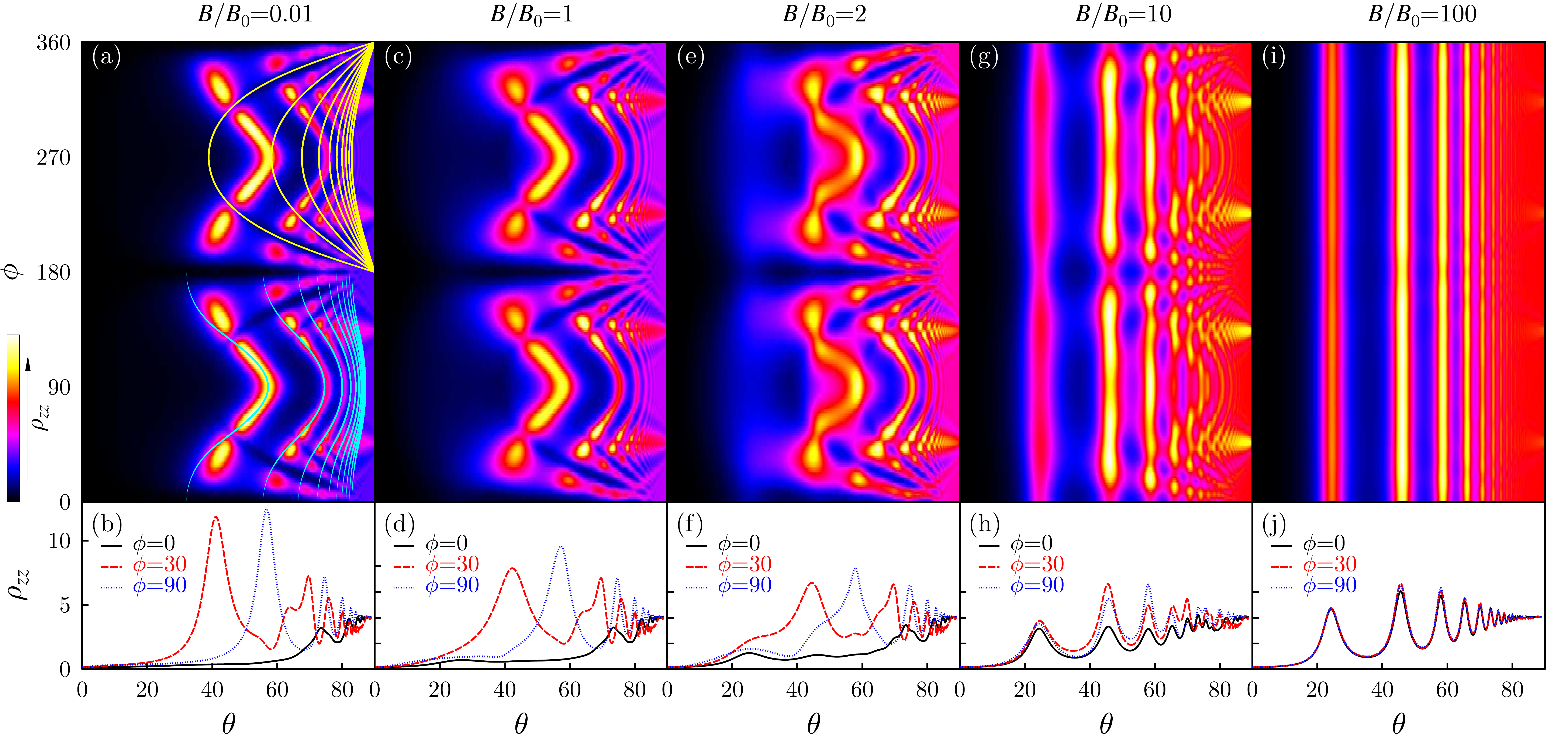}
\caption{Calculated resistivity as a function of $\phi$ and $\theta$
  for different values of $B/B_0$.  In these simulations, the
  parameters are chosen to conform approximately to those appropriate
  for experiments on \cuscn ($m^*=5m_{\rm e}$, $B=45$\,T,
  $\tau=3$\,ps), see \cite{bamroexp,Goddard2004}. 
The yellow lines in (a) show the expected minima for
  Lebed oscillations and the blue lines show the expected maxima for
  Yamaji oscillations.
\label{sim}}
\end{figure*}

Each component of the vectors $\lambdab_\pm$ and $\lambdab_{\rm init}$
contributes for a specific
segment of the orbit and the exponential factors multiplying some of
the components are present in order to cancel the initial
exponential damping of the integrand. This damping does not depend
on a specific segment but it depends on the length of trajectory
before reaching this specific segment. This exponential damping is 
taken into account in the vector ${\bf x}_0$. This vector 
dictates the evolution  of the
quasiparticle's path and includes all the processes at the MB junctions. 
The components of the vector ${\bf x}_n$ are as
follows:
\begin{eqnarray}
\alpha_{1}^{n}=\lambda^{\alpha_{1}}_{-}e^{-{\rm i}
n\Delta k_z}+ap\beta_{1}^{n}+aq\alpha_{2}^{n+1},\\
\beta_{1}^{n}=\lambda^{\beta_{1}}_{-}e^{-{\rm i}
n\Delta k_z}+bp\alpha_{2}^{n}+bq\beta_{1}^{n-1},\\
\alpha_{2}^{n}=\lambda^{\alpha_{2}}_{-}e^{-{\rm i}
n\Delta k_z}+ap\beta_{2}^{n}+aq\alpha_{1}^{n-1},\\
\beta_{2}^{n}=\lambda^{\beta_{2}}_{-}e^{-{\rm i}
n\Delta k_z}+bp\alpha_{1}^{n}+bq\beta_{2}^{n+1},
\end{eqnarray}
where $a=\exp(-2\xi/\omega\tau)$ and $b=\exp(-(\pi-2\xi)/\omega\tau)$
are the increments in the damping exponential after traversing an
$\alpha$ or $\beta$ segment respectively.  These recursive equations
encode all the information about the behaviour at the MB junctions.  Because
$x_{n\pm1}=x_{n}e^{\mp{\rm i}\Delta k_{z}}$ we can write them as a
single vector equation
\begin{equation}
{\bf x}_{n}=\lambdab_{-}e^{-{\rm i} n\Delta k_z}+\Gammam {\bf x}_{n},
\label{intermediate}
\end{equation}
where
${\bf
  x}_{n}=(\alpha_{1}^{n},\beta_{1}^{n},\alpha_{2}^{n},\beta_{2}^{n})$
and the matrix $\Gammam$ is given by a product of two matrices, one
describing damping and the other taking into account the connection
between orbit segments:
\begin{equation}
\Gammam=\left(
\begin{array}{cccc}
a & 0 & 0 & 0 \\
0 & b & 0 & 0 \\
0 & 0 & a & 0 \\
0 & 0 & 0 & b \\
\end{array} \right)\left(
\begin{array}{cccc}
0 & p & q e^{-{\rm i} \Delta k_z} & 0 \\
0 & q e^{{\rm i} \Delta k_z} & p & 0 \\
q e^{{\rm i} \Delta k_z} & 0 & 0 & p \\
p & 0 & 0 & q e^{-{\rm i} \Delta k_z} \\
\end{array} \right).
\end{equation}
Eq.~(\ref{intermediate}) is readily solved by assigning $n=0$ to the
FS slice initially occupied by the quasiparticle.  Thus with ${\bf
x}_{0}=(\Identity-\Gammam )^{-1}\lambdab_{-}$, where $I$ is $4\times
4$ identity matrix, we obtain
\begin{equation}
\sigma_{zz}=\lambdab_{+}\cdot\lambdab_{\textrm{init}}+\lambdab_{+}\cdot\Gammam(\Identity-\Gammam
)^{-1}\cdot\lambdab_{-},\label{conductivity}
\end{equation}
which is the main result of this paper.
The expression for the $\Gammam(\Identity-\Gammam )^{-1}$ matrix is:
\begin{eqnarray}
\Gammam(\Identity-\Gammam )^{-1}=\frac{1}{N}\left( \begin{array}{cccc}
t & ap {r}^* & a{r^*s^*} & a^{2}p {s^*} \\
abp s & w & bp {r^*} & abp^{2} \\
a rs & a^{2}p s & t & ap r \\
bp r & abp^{2} & abp {s}^* & {w}^* \\
\end{array} \right),
\end{eqnarray}
where
$N=1+b^{2}q^{2}-a^{2}(q^{2}+b^{2}(p^{2}-q^{2})^{2})
-2bq(1+a^{2}(p^{2}-q^{2}))\cos\Delta k_z$ and
\begin{eqnarray}
r&=&1-be^{{\rm i} \Delta k_z}q\\
s&=&e^{{\rm i} \Delta k_z}q+b(p^{2}-q^{2})\\
t&=&1+b^{2}q^{2}-2bq\cos \Delta k_z-N\\
w&=&b(qe^{{\rm i} \Delta k_z}{r^*}+a^{2}(p^{2}-q^{2})s).
\end{eqnarray}
The integrals in Eq.~(\ref{lambda}) can be evaluated using the
Jacobi-Anger expansion $e^{{\rm i}
z\cos\theta}=\sum_{k=-\infty}^{\infty}{\rm i}^{k}J_{k}(z)e^{{\rm i}
k\theta}$ where $J_{k}(z)$ is a $k$-th order Bessel function of the
first kind and hence $\int
E_{\pm}(\theta)d\theta=\sum_{k=-\infty}^{\infty} {\rm i}^{k}J_{k}(\pm
\eta) z_k^\pm e^{\theta({\rm i} k\pm 1/\omega\tau)}$, where
$z_{k}^{\pm}=e^{-{\rm i} k(\phi-\frac{\pi}{2}+\xi)}/({\rm i}
k\pm1/\omega\tau)$.  This implies that
\begin{equation}
\lambdab_{\pm} = \sum_{k=-\infty}^{\infty}{\rm i}^{k} z_{k}^{\pm}
\left(\begin{array}{c} 
J_{k}(\pm\eta) (e^{2{\rm i}k\xi}e^{\pm2\xi/\omega\tau}-1) \\
J_{k}(\pm\eta) ((-1)^k e^{\pm(\pi-2\xi)/\omega\tau}-e^{2{\rm i}k\xi}) \\
J_{k}(\mp\eta) (e^{2{\rm i}k\xi}e^{\pm2\xi/\omega\tau}-1) \\
J_{k}(\mp\eta) ((-1)^k e^{\pm(\pi-2\xi)/\omega\tau}-e^{2{\rm i}k\xi})
\end{array}\right),
\label{integrals}
\end{equation}
and similar techniques can be used to evaluate $\lambdab_{\rm init}$.
Our results reduce to the expressions given by Yagi {\sl et al.}
\cite{Yagi1990} when $p\to 1$ or $\xi\to\frac{\pi}{2}$ while we can
extract the expressions for Lebed magic angles \cite{lebed} and Yamaji
maxima \cite{Yamaji1989} when $p \to 0$.  To study the general case,
we have encoded the solution in a computer program, separating each
term contributing to Eq.~(\ref{conductivity}) into real and imaginary
parts, and have summed up the Bessel functions in
Eq.~(\ref{integrals}), truncating the series at small enough $J_k$;
this involves typically about 200 Bessel functions in the sum.

\begin{figure}
\includegraphics[width=7.5cm]{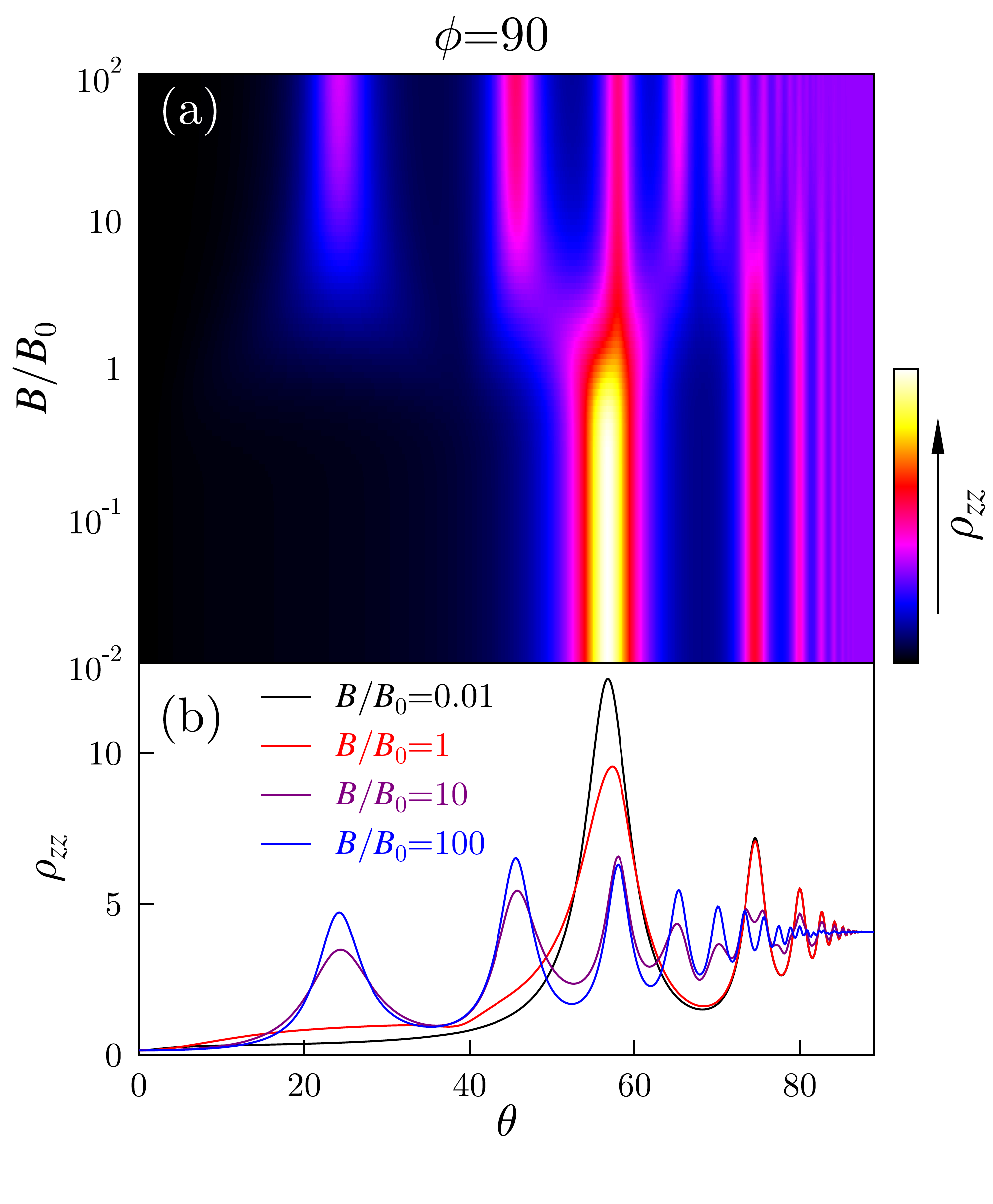}
\caption{Calculated resistivity as a function of $B/B_0$ for
$\phi=90^\circ$ for the same parameters as in Fig.~\ref{sim}.
\label{sim2}}
\end{figure}

In Fig.~\ref{sim} we show the calculated resistivity as a function of
$\phi$ and $\theta$ for different values of $B/B_0$ using parameters
chosen to conform approximately to those appropriate for experiments
on \cuscn.  In these calculations $B$, $\tau$, and hence $\omega_{\rm
c}\tau$, are fixed and only $B_0$ is varied.  When $B_0$ is very large
(small $B/B_0$), no MB occurs and the calculated $\rho_{zz}$ shown in
Fig.~\ref{sim}(a,b) display a rich structure comprising features
originating from both open and closed orbits.  Some of the minima are
due to Lebed oscillations \cite{lebed} resulting from the open orbits
[see Fig.~\ref{fs}(b)] and their expected angular dependence is shown
in the yellow lines in the upper part of Fig.~\ref{sim}(a) and is
given by $\tan\theta = {n \pi / k_{\rm F}\cos\xi\sin\phi}$, for
integer $n\geq 1$ \cite{lebed}.  Some of the maxima are due to Yamaji
oscillations \cite{Yamaji1989,Yagi1990,McKenzie1998} resulting from
the lens-like closed orbits [see Fig.~\ref{fs}(c)] as shown in the
blue lines in the lower part of Fig.~\ref{sim}(a).  The angular
positions of these are known to be given by a ``caliper'' measurement
of the FS \cite{caliper} which a simple calculation can translate into
the present geometry as
\begin{equation}
\tan\theta \approx {\pi (n-\frac{1}{4}) \over k_{\rm F}} \times
\left\{ \begin{matrix} (1-\sin\phi\cos\xi)^{-1} &
  \vert\phi-\frac{\pi}{2}\vert<\xi \cr
(\cos\phi\sin\xi)^{-1} & \mbox{otherwise.} 
   \end{matrix}
\right.
\end{equation}
The values of $\phi$ at which either the Lebed or Yamaji oscillations
dominate the calculated resistivity in Fig.~\ref{sim} are similar to
those found experimentally in \cuscn \cite{Goddard2004}.

As $B_0$ decreases this structure begins to break up as MB 
starts to occur and in the limit of low $B_0$
(see Fig.~\ref{sim}(i,j)) the only dominant orbits are 
breakdown orbits (see Fig.~\ref{fs}(d)) leading to the observation of
$\phi$-independent Yamaji oscillations with maxima at
$\tan\theta\approx \pi(n-\frac{1}{4})/k_{\rm F}$.  Because the
breakdown probability is ${\rm e}^{-B_0/B\cos\theta}$, it is
noticeable [particularly in Figs.~\ref{sim}(h) and \ref{sim}(j)] that
$\rho_{zz}$ is more $\phi$-independent at low $\theta$ than at high
$\theta$ since MB becomes less likely as $\theta$ increases.  We note
that at low-fields it is also possible to observe Danner-Chaikin-like
oscillations \cite{Danner1994} from the open sections close to
$\phi=0$ and $\theta=\frac{\pi}{2}$.

The transition between the low-field and high-field behavior can be
studied by fixing $\phi$ and varying $B/B_0$ and this is shown
in Fig.~\ref{sim2}.  The series of Yamaji maxima (starting at
$\theta\approx 55^\circ$), which dominates the response at low $B/B_0$
due to the small lens-like orbit in Fig.~\ref{fs}(c), give way at high
$B/B_0$ to a different series of Yamaji maxima (starting at
$\theta\approx 23^\circ$) due to BAMROs resulting from the breakdown
orbit in Fig.~\ref{fs}(d).  The crossover between the two regimes
begins above $B/B_0\approx 1$ which is when the probability of MB
becomes significant.

There is a remarkable similarity between the predictions of this model
and the data of Ref.~\onlinecite{bamroexp,Goddard2004}.  It is
expected that this approach to summing all the contributions to the MB
network model will open up new avenues in research on low-dimensional
metals.

We thank EPSRC (AN) and the Oxford University Glasstone Fund (PAG)
for financial support.

\end{document}